\documentclass[twocolumn,amsmath,amssymb]{revtex4}
\usepackage{graphicx}
\usepackage{hyperref}
\begin{document}

\title{Binary Inspirals in Nordstr\"{o}m's Second Theory}

\date{\today}

\author{Travis M. Garrett}
\email{tgarrett@perimeterinstitute.ca}
\affiliation{Perimeter Institute for Theoretical Physics, Waterloo, Ontario N2L 2Y5, Canada}


\begin{abstract}
We investigate Nordstr\"{o}m's second theory of gravitation, 
with a focus on utilizing it as a testbed for developing techniques
in numerical relativity.
Numerical simulations of inspiraling compact star binaries
are performed for this theory,  
and compared to the predictions of semi-analytic calculations  
(which are similar to Peters and Mathews' results for GR).
The simulations are based on a co-rotating spherical coordinate system, 
where both finite difference and pseudo-spectral methods are used. 
We also adopt the ``Hydro without Hydro" approximation, and 
the Weak Radiation Reaction approximation when the orbital motion is quasi-circular.
We evolve a binary with quasi-circular initial data for hundreds of orbits and find that 
the resulting inspiral closely matches  
the 1/4 power law profile given by the semi-analytical calculations.  
We additionally find that an eccentric binary 
circularizes and precesses at the expected rates.
The methods investigated thus provide a promising 
line of attack for the numerical modeling of long binary inspirals in general relativity.
\end{abstract}

\maketitle

\newcommand{\be}{\begin{equation}}
\newcommand{\ee}{\end{equation}}

\section{Introduction}

Numerical methods capable of modeling the strong gravitational 
fields of merging black holes and neutron stars 
were found several years ago after many years 
of searching.  
The main techniques are centered around the 
BSSN (see \cite{Nakamura:1987zz,Shibata:1995we,Baumgarte:1998te,clmz05,bcckm05}) 
and Generalized Harmonic (GH)
formulations \cite{Garfinkle:2001ni,pre05} of general relativity.  
These two classes of techniques generally 
agree in their predictions \cite{baker07},\cite{Hannam:2009rd}, and also match the  
Post-Newtonian calculations during the inspiral \cite{bmmck06}.
We are designing our own numerical relativity techniques 
for comparable mass binaries with the goal of modeling long 
stretches of the inspiral phase preceding the merger  
(as could be useful in discriminating 
neutron star equations of state \cite{Read:2009yp}).  
We employ corotating spherical coordinate systems to this end, 
as they are well adapted to the slow evolution of the binary during this phase. 
This is similar to the approach of the Caltech group (\cite{scheel06,scheel08,Szilagyi:2009qz}) 
although we differ in most of the details.  

Spherical coordinate systems have several attractive features from a 
computational standpoint.  A spherical mesh has a high 
density of grid points near the origin where it is needed to resolve 
the high curvature potential wells of the compact bodies.
This thus provides an alternative to the Adaptive Mesh Refinement 
methods (see e.g. \cite{Berger:1984zza},\cite{lehner05}) which have seen success lately.   
Additionally, a spherical grid provides a $S^2$ outer boundary, 
which is more suited to outgoing waves than a cubic grid.
One of the key features is that it allows for the use of pseudo-spectral 
methods: by using spherical harmonics the 3+1 problem can 
be split into a set of 1+1 problems which can be solved with fast 
and accurate implicit methods.  
The usage of spherical harmonics likewise avoids the complications
due to the coordinate singularities in $\theta$ and $\phi$. 
A spherical system also meshes nicely 
with corotating coordinates, 
as proposed in \cite{brady98}.
This removes most of the dynamical motion in the grid, 
thus helping to cut down on spurious field excitations.

We are currently supplementing this primary numerical method with 
some useful additional approximations.  
Neutron star like compact bodies are our focus, 
and thus the complications of modeling black holes 
are avoided.
Furthermore we make use of the ``Hydro-without-Hydro" 
approximation \cite{bhs99} so that a relativistic 
fluid dynamics code is not currently needed.
Radial profiles for isolated polytropic stars are generated 
and then inserted into a binary system. 
The gravitational fields and net accelerations of the stars
in the binary are evolved,
while the radial density profiles of the stars are held constant.
This is a good approximation as geodesic deviations due to 
tidal distortions of the stars only arise at high
post-Newtonian order.

Another useful approximation is possible when the stars are in 
quasi-circular orbit (which is generally the case for 
astrophysical binaries as the gravitational radiation circularizes the orbit).
As pointed out in \cite{brady98}, in a corotating coordinate system 
a quasi-circular binary evolves at the slow radiation reaction time scale 
$\tau \sim d^4/M^3$ for a separation $d$ and mass $M$.
This makes the Weak Radiation Reaction (WRR) approximation
possible: the second time derivatives in the field equations can be dropped, thus 
changing the character of the differential equations.

There are some numerical drawbacks to using spherical 
coordinate systems.  
A primary one is that 
spherical harmonic decomposition 
and synthesis are computationally expensive.
For instance the decomposition:
\be
A_{lm}(r) = \int f(r,\theta,\phi) Y^{\ast}_{lm}(\theta,\phi) d\Omega
\ee
is an $\mathcal{O}(N^5)$ calculation if  
discretized in a direct manner,  
although this can be reduced to $\mathcal{O}(N^4)$
 by defining intermediate variables and splitting the 
integration into two steps.  
The computational cost is further reduced 
in our case as we only need decompose and synthesize in the small 
volumes containing our compact bodies.  
This makes it possible to run 
high resolution meshes in a reasonable amount of time on a single computer.
Further reduction the the order of these operations is also 
possible: \cite{driscoll94,alpert91,healy03}, 
although these methods have a large amount of overhead 
and only become efficient for large values of $l$ and $m$.

There are some other subtle issues that arise in spherical coordinate systems.
For instance when using second order accurate methods 
one finds that very fine meshes are needed to accurately 
compute the accelerations of the stars.  
We find that using 
higher order methods allows for accurate results with much lower 
resolution.  This is described in greater detail in section 3.

Before we attempt to use these techniques in a full GR code we have 
elected to test them in a scalar gravity model theory 
(following the advice of \cite{shapiro93} and \cite{watt99}).
This forms the focus of this paper.
After the tests are successful 
we can then feel confident in applying the methods  
to a full general relativistic code.  
After considering several scalar theories, 
we chose Nordstr\"{o}m's second theory (see \cite{ni72}).
It is a fully conservative metric 
theory, with parameterized post-Newtonian parameters 
$\gamma=-1$, $\beta=1/2$ and all others zero.
The theory thus has no Nordtvedt effect (see \cite{nordtvedt68}), and the stars
move on Keplerian orbits in the limit of small mutual gravitational potential.  
This is nontrivial since the Nordtvedt effect
can produce considerable deviation from Keplerian orbits for highly compact 
bodies, even at arbitrarily large separations.  
We can therefore perform a semi-analytic calculation of the 
rate of energy and angular momentum loss for a binary 
based on the Keplerian motion, in a similar fashion to 
Peters and Mathews' calculation in GR (\cite{peters63},\cite{peters64}).
We find that the radiation is quadrupole to the leading order, 
despite being based on scalar field, due to approximate 
conservation of mass and momentum.

The fully conservative nature of Nordstr\"{o}m's theory has an additional
benefit in that it justifies the use of the ``Hydro-without-Hydro" approximation.  
The approximation is safe since corrections to the star's equilibrium do not show up through  
first post-Newtonian order for fully conservative theories \cite{wiseman97}.

In the following sections we first review the details of Nordstr\"{o}m's second theory.  
We describe the construction of isolated stars in this theory, 
and the equations of motion for the matter. 
We then review the 1PN equations of motion, which lead to 
precession for eccentric binaries.
We calculate the energy and angular momentum loss on a 
large $S^2$ in the wave zone, and apply the results to a Keplerian system,  
which cause the binary to decay and circularize in a 
similar fashion to GR.
Next we detail the numerical methods used to simulate the binary, 
including switching to a co-rotating reference frame, 
and several preliminary test cases that are needed to check the performance of the code.
We finish with the results of simulating Nordstr\"om binaries in both 
quasi-circular and slightly eccentric orbits.  

We obtain a nice agreement 
between our simulation and the analytic calculations.
The orbital decays of quasi-circular and eccentric binaries closely 
match the expectations from our semi-analytic calculations, 
as seen in the evolution of the semi-major axis 
$a$ and eccentricity $e$.
We also find that the eccentric binaries precess at the rate predicted 
by 1st Post Newtonian calculations, which gives supporting 
evidence that Nordstr\"om's theory obeys the Strong 
Equivalence Principle (SEP).

\section{Nordstr\"{o}m's Second Theory}

After the development of special relativity it became clear that Newton's 
gravitational theory could no longer be completely correct.
For instance the Poisson equation is solved simultaneously throughout all of space, 
and thus the acceleration of a massive body would allow for the instantaneous 
transmission of signals, in disagreement with the finite speed of light.    
Comparisons with electromagnetic theory suggested that the 
spatial Laplacian operator should be replaced with the D'Alembertian wave operator.  
Gunnar Nordstr\"{o}m used this idea to develop two relativistic scalar gravity 
theories a couple years before Einstein discovered General Relativity  
(we focus on the second, more complete theory).  
Nordstr\"{o}m's theory is not a viable candidate for describing 
relativistic gravitation, as Nordstr\"{o}m quickly 
realized after Einstein presented his tensor theory.  For instance, being 
conformally flat, it predicts zero bending of light.  However, it does 
provide an excellent arena for developing tools for numerical relativity.

Einstein and Fokker demonstrated that Nordstr\"{o}m's second theory
could be expressed in geometric terms.  The metric is conformally flat 
(thus the Weyl tensor is zero: $C_{\alpha\beta\gamma\delta}=0$) and is 
specified by a scalar field $\varphi$:
\be
g_{\mu\nu} = (1+\varphi)^2 \eta_{\mu\nu}.
\label{metric}
\ee
The scalar field is generated in turn by setting the 
Ricci scalar to be equal to the trace of the 
stress energy tensor (see \cite{ni72}):
\be
R = 24 \pi T.
\label{eomfield0}
\ee
For our simulations we will use
 the standard perfect fluid stress energy tensor:
\be
T^{\mu\nu} = (\rho + \rho\varepsilon + p)u^{\mu} u^{\nu} + g^{\mu\nu} p,
\ee
so the field equation becomes:
\be
\Box \varphi = 4 \pi (1+\varphi)^3 (\rho + \rho\varepsilon -3 p).
\label{eomfield}
\ee

Being conformally flat, Nordstr\"{o}m's theory effectively 
has a background geometry $\eta_{\mu\nu}$,
and a conserved energy-momentum complex can be constructed:
\begin{align}
t^{\mu\nu} = \frac{1}{4\pi} \left[ \eta^{\mu\alpha}\eta^{\nu\beta}\varphi_{,\alpha}\varphi_{,\beta}
-\frac{1}{2} \eta^{\mu\nu}\eta^{\alpha\beta}\varphi_{,\alpha}\varphi_{,\beta} \right]  
\label{energy} \\ \notag +(1+\varphi)^6 T^{\mu\nu},
\end{align}
with ${t^{\mu\nu}}_{,\nu}=0$.
Nordstr\"{o}m's theory has gravitational waves, although they are different 
from those in general relativity since $C_{\alpha\beta\gamma\delta}=0$.
Namely, the waves occur in the scalar field $\varphi$ as described by equation (\ref{eomfield}), 
and (\ref{energy}) describes their localizable energy density: 
$t^{00}= (1/8\pi)[(\partial_t\varphi)^2 +(\nabla\varphi)^2]$.

\subsection{Single Star Solution}

We utilize the hydro-without-hydro approximation in our code.  
This is a good approximation since 
tidal distortions of the stars do not affect the orbital dynamics 
until relatively high Post-Newtonian order.  We thus need to construct a model of 
a single star in Nordstr\"{o}m's theory to use as a source in the main code.  For a single
isolated star equation (\ref{eomfield}) becomes:
\be
\frac{1}{r^2}\partial_r(r^2\partial_r \varphi) = 4\pi (1+\varphi)^3 [\rho(1+\varepsilon)-3p].
\label{sseos1}
\ee
We will assume the star is a polytrope, so that the pressure and internal energy are given by:
\begin{align}
p=\kappa \rho^{\Gamma}, \\
\varepsilon = \frac{\kappa}{\Gamma-1} \rho^{\Gamma-1},
\end{align}
which reduces equation (\ref{sseos1}) to two variables.  
Application of the projection operator
$Q^{\alpha}_{\mu} = u^{\alpha}u_{\mu}+\delta^{\alpha}_{\mu}$ on the divergence 
of the stress energy tensor ${T^{\mu\nu}}_{;\mu}=0$ gives: 
\be
\rho h u^{\nu} {u^{\alpha}}_{;\nu} + Q^{\alpha\nu} p_{,\nu} = 0
\label{sseos3}
\ee
where $h$ is the relativistic specific 
enthalpy: $h=h(\rho)=1+\varepsilon +p/\rho$.  
With $u^0 = 1/(1+\varphi)$ 
and $u^i=0$ this reduces this to:
\be
\frac{d\varphi}{dr} = -\frac{(1+\varphi)}{h}\frac{dh}{dr}.
\label{sseos2}
\ee

We solve (\ref{sseos1}) and (\ref{sseos2}) iteratively.  
In the first iteration the 
$(1+\varphi)^3$ and $(1+\varphi)$ terms are dropped respectively, 
resulting in a modified Lane-Emden equation 
which is solved by specifying the central pressure $\rho_c$ 
and integrating out.  
For successive iterations the $(1+\varphi)^3$ 
and $(1+\varphi)$ terms are reinserted 
with the previous iteration's solution for $\varphi$.  
The iterative process converges 
for values of $\rho_c$ below some critical value 
(which depends on $\kappa$ and $\Gamma$). 
For instance, with a stiff equation of state with $\Gamma=2$ 
we find the most compact star that can be formed has radius $R \sim 1.75M$, 
which is smaller than the event horizon radius in classical GR 
(in addition to being smaller than the Buchdahl-Bondi bound of $R=9/4M$ for GR).
We use the softer value of $\Gamma=5/3$ to generate stars for the simulation, 
as the smoother transition to zero density at the outer boundary of the star 
allows the spherical harmonic decomposition to converge more quickly.
 
\subsection{Matter Equations of Motion}

The isolated star solutions will be  
used in the binary simulation 
with the hydro-without-hydro approximation, 
so that only the net accelerations of the centers of masses need to be found.  
The conserved density $\rho^{\ast}$ (see \cite{tegp}) proves useful to this end:  
\be
\rho^{\ast} = \rho (-g)^{1/2} u^0 = \rho \frac{(1+\varphi)^3}{\sqrt{1-v^2}},
\label{rhostar}
\ee
with $\partial_t \rho^{\ast} + \partial_i (\rho^{\ast} v^i) = 0$.
 The conserved density can be used to define the center of mass for star $a$:
 \be
 x^j_a =  \frac{1}{M^{\ast}_a} \int_a \rho^{\ast} x^j d^3x ,
 \ee
 where ${M^{\ast}_a}$ is volume integral of $\rho^{\ast}$.
Differentiating twice gives the coordinate acceleration $a^i_a$ of the 
center of mass:
\be
a^j_a = \frac{1}{M^{\ast}_a} \int_a \rho^{\ast} \frac{d v^j}{dt} d^3x.
\label{bseom2}
\ee
An expression for the local accelerations $d v^j/dt$ 
is then found by
expanding the projection of the divergence of the stress energy tensor (\ref{sseos3}):
 \begin{align}
 \rho^{\ast} \frac{d v^j}{dt} = - \rho^{\ast} \Gamma^j_{\alpha \beta} v^{\alpha} v^{\beta} 
 -(u^0)^{-1} \rho^{\ast} v^j  \frac{d u^0}{dt} \label{bseom3} \notag \\ 
  -(u^0)^{-2} \frac{\rho^{\ast}}{\rho h}Q^{j\nu}p_{,\nu} .
 \end{align}
 During the simulation the terms in (\ref{bseom3}) are calculated and 
 integrated with (\ref{bseom2}) to find the net acceleration of the star 
 (with the radial profile for $\rho^{\ast}$ kept constant).
 
\subsection{Semi-Analytical Calculation of Orbital Evolution}

Semi-analytic calculations of the evolution of a binary 
system in N\"ordstrom's theory are needed so that comparisons can 
be made with the numerical simulations.  
We first review the 1PN results for the theory, which follow 
from having the parameterized post-Newtonian parameters 
given by $\gamma=-1$, $\beta=1/2$ (and all others zero).
The theory therefore has no Nordtvedt effect 
(since $4\beta - \gamma - 3 = 0$ -- see \cite{tegp})
and compact stars move on Keplerian orbits at large separations.
The 1PN analysis also leads to precession for 
eccentric binaries.  Next we will solve for the 
energy and angular momentum radiated at large radius 
by a binary on a Keplerian orbit, and then 
apply these results to evolve the orbital parameters.

Consider first a binary with stars $a$ and $b$ in a slightly eccentric orbit
with an angular velocity $\omega$.
Arrange a Cartesian coordinate system so that the binary is instantaneously 
aligned along the x-axis, 
with a separation $d$ and the bulk of the velocity in the $\hat{y}$ direction: 
$v^y_a \sim \omega d/2$, and a small amount of radial velocity $v^x_a$.
The PN analysis then finds that the acceleration $\ddot{x}_a$ of star $a$ 
in the $\hat{x}$ direction is:
\begin{align}
\ddot{x}_a = -\frac{M_b}{d^2} \left(1+\frac{M_b}{d}-v^2_a \label{ddotx}
- \frac{3}{2} (v^x_b)^2 - (v^x_a - v^x_b)v^x_b \right) ,
\end{align}
 where 
$M_b = \int_b \rho^{\ast}(1+(1/2)\bar{v}^2-(1/2)\bar{U}+\varepsilon) d^3x$
 is the gravitational mass (equal to the inertial mass) of star $b$ 
 (with $\bar{U}=\int_b \rho' |\bold{x}-\bold{x}'|^{-1} d^3x'$ 
 and $\bar{\bold{v}}=\bold{v}-\bold{v}_0$ where $\bold{v}_0$ is velocity of the 
 center of mass).   
The acceleration in the $\hat{y}$ direction is:
\begin{align}
\ddot{y}_a = \frac{M_b (v^y_a-v^y_b) v^x_b}{d^2} \label{ddoty} .
\end{align}
  
 At large separations the post-Newtonian corrections 
 $M_b/d$ and $v^2_a$ drop out, leaving the binary in a Keplerian orbit.
This confirms that Nordstr\"{o}m's second theory 
has no Nordtvedt effect through 1PN order (and as a purely 
geometric theory it should follow the SEP).
One further consequence of the post-Newtonian corrections 
(\ref{ddotx}) and (\ref{ddoty}) is that an eccentric binary will 
precess.  A binary with semi-major axis 
$a$, eccentricity $e$, and total mass $M$ 
will precess $\Delta \tilde{\omega}$ radians 
per orbit \cite{tegp}:
\be
\label{precess_an}
\Delta \tilde{\omega} = -\frac{\pi M}{a (1-e^2)} .
\ee 
This is six times slower and in the opposite direction as GR.

\subsection{Energy Loss at Outer Boundary}

We next find the energy radiated by a binary due to the waves
generated in $\varphi$.
The energy-momentum complex
\ref{energy} follows the standard flat-space conservation law:
\be
{t^{\mu\nu}}_{,\nu}=0. \label{cons1}
\ee
Integration inside a volume bounded by $S^2$ gives:
\be
\int {t^{00}}_{,0} d^3 x = - \int {t^{0i}}_{,i} d^3 x,
\ee
or, using Gauss's law:
\be
\partial_t E = - \int t^{0i} n_i dS.
\ee
with $t^{0i}$ given by:
\be
t^{0i} = -\frac{1}{4\pi}\partial_t \varphi \partial_i \varphi.
\ee
The spatial derivative 
$\partial_i \varphi$ can be transformed into $-n^i \partial_t \varphi$ since at large 
radius $\varphi$ is approximately a spherical wave: $\varphi \sim \sin(t-r)/r$.
This leads to:
\be
\partial_t E = - \frac{1}{4\pi} \int (\partial_t \varphi)^2 r^2 d\Omega 
\label{eloss1}
\ee
for the energy loss.

An expression for $\partial_t \varphi$ is needed next.  Rewriting the equation 
of motion for the field (\ref{eomfield}) with the conserved mass density $\rho^{\ast}$ gives:
\be
\Box \varphi = 4 \pi \rho^{\ast} (1-v^2)^{1/2} (1 + \varepsilon -3 p/\rho).
\label{eomfield2}
\ee
This can be solved with a Green's function:
\be
\varphi = -\int \frac{[{\rho^{\ast}}'(1-(1/2)v'^2+\varepsilon' - 3p'/\rho' +\mathcal{O}(v^4))]_{ret}}
{|\bold{x}-\bold{x}'|} d^3 x' \label{green1},
\ee
which is evaluated at the retarded time $t'=t-|\bold{x}-\bold{x}'|$.
The $1/|\bold{x}-\bold{x}'|$ term expands into:
\be
\frac{1}{|\bold{x}-\bold{x}'|} = \frac{1}{r} + \frac{x^j {x^j}'}{r^3} + \dots
\ee
Only the first term is needed since it is utilized at large radius.  
Likewise only the first term in the expansion:
\be
r-|\bold{x}-\bold{x}'| = \frac{x^j {x^j}'}{r} 
+ \frac{x^j x^k}{2r}\frac{({x^j}'{x^k}'-{r'}^2\delta^j_k)}{r^2} + \dots
\ee
will be needed.
Equation (\ref{green1}) can thereby be expanded out in multipole moments:
\be
\varphi = - \frac{1}{r} \left[ M + n_i \partial_t D^i + \frac{1}{2} n_i n_j \partial^2_t Q^{ij} + \dots \right]
\label{varphi1}
\ee
with
\be
M = \int \rho^{\ast}{}'(1-(1/2)v'^2+\varepsilon' - 3p'/\rho' +\mathcal{O}(v^4)) d^3 x' ,
\ee 
\be
D^i =  \int \rho^{\ast}{}'(1-(1/2)v'^2+\varepsilon' - 3p'/\rho' +\mathcal{O}(v^4)) x^i{}' d^3 x' ,
\ee
and
\be
Q^{ij} =  \int \rho^{\ast}{}'(1-(1/2)v'^2+\varepsilon' - 3p'/\rho' +\mathcal{O}(v^4)) x^i{}' x^j{}' d^3 x' 
\ee
(and higher multipoles unnecessary for a leading order calculation of the energy loss).

The time derivatives of these multipole moments are needed for (\ref{eloss1}).
The quadrupole contribution $\partial^3_t Q^{ij}$ scales as $v^5$, 
so all terms of order $v^6$ and higher will be dropped for this leading order calculation.  
We first examine the time derivative of the monopole: 
\be
\partial_t M = \int \rho^{\ast}{}' \frac{d}{dt}(1-(1/2)v'^2+\varepsilon' - 3p'/\rho') d^3 x' , \label{ptmonopole}
\ee
where $(d/dt) \int_V  \rho^{\ast} f d^3 x = \int_V  \rho^{\ast} (df/dt) d^3 x$
has been used to pass the time derivative through $\rho^{\ast}$.
Note that if the two stars are on a quasi-circular orbit then the time derivatives 
in the integrand are on the radiation reaction timescale, and thus contribute 
radiation at a far lower scale than $v^5$.  
However, they do contribute to eccentric 
binaries where the separation oscillates on the orbital timescale.

First consider the $d\varepsilon/dt$ term in (\ref{ptmonopole}), and to simplify 
assume that $\Gamma=2$, although the result holds in general.  
The specific internal energy can be expanded: 
$\varepsilon=\kappa\rho=\kappa\rho^{\ast}(1-(1/2)v^2-3\varphi+\mathcal{O}(v^4))$, 
where only the first term can contribute through quadrupole order.
This term reduces to $d\varepsilon/dt=-\kappa\rho^{\ast}\partial_i v^i$.  
Thus any monopole radiation of order $v^5$ from the $d\varepsilon/dt$ 
term stems from the ``breathing" motion as the star 
expands and contracts during the elliptical orbit.  
However, Nordstr\"{o}m's second 
theory is fully conservative, and accordingly the stellar matter undergoes no ``breathing" motion:
the central density is constant to 1PN order 
(which justifies the use of the ``Hydro-without-Hydro" assumption).
Therefore the entire $d\varepsilon/dt$ term does not contribute at $v^5$ order.  
The same holds for the $d(3p/\rho)/dt$ term.  
We find the monopole contribution to the radiation:
\be
\partial_t M = -\frac{1}{2} \int \rho^{\ast}{}' \partial_t (v'^2) d^3 x' + \mathcal{O}(v^7),
\ee
which is at $v^5$ order if the orbit is eccentric and is much smaller otherwise
(note also that the total time derivative has been switched to a partial derivative 
since the velocity is now essentially constant throughout the star).   
Note that it is also convenient that radial pulsations do not contribute at leading order 
as this allows the stars to be treated as point bodies in later calculations.

We next examine the contribution from the dipole via $\partial^2_t D^j$.  
Distributing the two time derivatives through the integrand gives:
\begin{align}
\partial^2_t D^j = \int \rho^{\ast}{}' [(1-v'^2+\varepsilon'-3p'/\rho')  \frac{dv^j{}'}{dt} \label{dipole1} \\ \notag
- \frac{d}{dt}\left( \frac{dv^i{}'}{dt} v^i{}' \right)x^j{}' - 2 \frac{dv^i{}'}{dt} v^i{}' v^j{}' ] d^3 x' .
\end{align}   
The $\int \rho^{\ast}{}' (dv^j{}'/dt) d^3 x'$ term 
integrates to zero due to Newton's third law.  
Any corrections
would enter at $v^6$ order at the soonest, 
and all the other terms in equation (\ref{dipole1})
also scale as $v^6$.
Thus the dipole does not radiate to leading order:
\be
\partial^2_t D^j = 0 + \mathcal{O}(v^6).
\ee

The final multipole moment that can 
contribute through leading order is the quadrupole.  
The $(1/2)v'^2$, $\varepsilon'$, and $- 3p'/\rho'$ terms all enter 
at $\mathcal{O}(v^7)$, leaving only: 
\be
\partial^3_t Q^{ij} = \partial^3_t \int \rho^{\ast}{}' x^i{}' x^j{}' d^3 x'
+ \mathcal{O}(v^7).
\ee
   
Combining the terms from the monopole and quadrupole gives:
\be
\partial_t E = -\frac{1}{4\pi} \int (\partial_t M + (1/2)n_i n_j \partial^3_t Q^{ij})^2 d\Omega
\ee   
 for the rate of energy loss.  With the integrals:
 \begin{align}
 \int n_i n_j d\Omega = \frac{4\pi}{3}\delta^{ij} \label{s2int}, \\ \notag
 \int n_i n_j n_k n_l d\Omega = \frac{4\pi}{15}[\delta^i_j\delta^k_l+\delta^i_k\delta^j_l+\delta^i_l\delta^j_k] 
 \end{align} 
 this becomes:
 \be
 \partial_t E = -(\partial_t M)^2 -\frac{1}{3}\partial_t M \partial^3_t Q^{ii}
 -\frac{1}{60} (\partial^3_t Q^{ii})^2 - \frac{1}{30} (\partial^3_t Q^{ij})^2. \label{edot1}
 \ee
 This reduces to:
 \be
 \partial_t E =  - \frac{1}{30} (\partial^3_t Q^{ij})^2
 \ee
 for zero eccentricity case, which is six times smaller than the value found in GR.
 
\subsection{Angular Momentum Loss at Outer Boundary}

A similar calculation determines the rate at which the angular momentum 
is radiated.  With equation (\ref{cons1}) and:
 \be
 L_i  = \epsilon_{ijk} \int x^j t^{0k} d^3x
 \ee
 one finds: 
  \be
 \frac{d L_i}{dt} = - \epsilon_{ijk} \int x^j {t^{kl}}_{,l} d^3 x = - \epsilon_{ijk} \int x^j t^{kl} n^l dS.
 \ee
The spatial stress energy tensor components are: 
\be
t^{kl} = \frac{1}{4\pi} \partial_k \varphi \partial_l \varphi 
- \frac{1}{8\pi} \delta^k_l \eta^{\alpha\beta} \partial_{\alpha} \varphi \partial_{\beta} \varphi,
\ee
and thus:
  \be
 \frac{d L_i}{dt} = - \frac{1}{4\pi} \epsilon_{ijk} \int x^j \partial_k \varphi \partial_l \varphi n^l dS.
 \ee
 The $ \eta^{\alpha\beta} \partial_{\alpha} \varphi \partial_{\beta} \varphi$ term is zero 
due to the anti-symmetry of  $\epsilon_{ijk}$.

Only one spatial partial derivative should 
both converted into a time derivative via 
$\partial_i \varphi = - n^i \partial_t \varphi$:
\be
 \frac{d L_i}{dt} = - \frac{1}{4\pi} \epsilon_{ijk} \int x^j \partial_k \varphi 
 \frac{1}{r}(\partial_t M + (1/2)n_l n_m \partial^3_t Q^{lm}) dS,
 \label{ldot1}
\ee
as converting both yields zero due to 
the anti-symmetry of $\epsilon_{ijk}$
(note also that the $n_l \partial^2_t D^l$ term has been 
dropped from $\partial_t \varphi$ since it doesn't 
contribute at leading order).  
In particular, examination of $\epsilon_{ijk}$ shows 
that the $\partial_k \varphi$ term that needs to be expanded to higher order,
while the $\partial_l \varphi$ can approximated by the time derivative.  
Application of the spatial derivative $\partial_k$ 
to the monopole term in (\ref{varphi1}) gives:
\be
\partial_k \left( -\frac{1}{r} M \right) = \frac{x^k}{r^3} M.
\ee
Insertion of this term into (\ref{ldot1}) also gives zero due to 
the anti-symmetry of $\epsilon_{ijk}$, 
and the same applies to all terms stemming from 
derivatives of powers of $r$.  
The only remaining relevant terms in $\partial_k \varphi$ are:
\be
\partial_k \varphi = - \frac{1}{r^2} \partial_t D^k - \frac{x^l}{r^3}\partial^2_t Q^{lk} + ...
\ee
Upon expansion in (\ref{ldot1}) 
the $- \partial_t D^k/r^2$ terms are multiplied by either one or three $n^i$ terms, 
and as $\int n_i d\Omega = \int n_i n_j n_k d\Omega= 0$  
these dipole terms integrate to zero.

Equation (\ref{ldot1}) thus becomes:
\be
 \frac{d L_i}{dt} = \frac{1}{4\pi} \epsilon_{ijk} \int n^j n^l \partial^2_t Q^{lk}
 (\partial_t M + (1/2)n_a n_b \partial^3_t Q^{ab}) d\Omega.
 \label{ldot2}
\ee
Reusing the integrals (\ref{s2int}) leads to:
\be
 \frac{d L_i}{dt} = \frac{1}{15} \epsilon_{ijk} \partial^3_t Q^{jl} \partial^2_t Q^{lk}
 \label{ldot3}
\ee
where the $\partial_t M$ term has also dropped out, again due to $\epsilon_{ijk}$.  
This expression is precisely one sixth of the value given by general relativity.

\subsection{Application to Keplerian Orbits}

We can now apply the equations for the rates of energy (\ref{edot1}) and angular momentum 
loss (\ref{ldot3}) to a binary star system in Keplerian orbit.  The stars have gravitational 
masses $m_1$ and $m_2$, a semi-major axis $a$ and eccentricity $e$.  The 
separation $d$ between the two stars is determined by the phase $\phi$:
\begin{equation}
d=\frac{a(1-e^2)}{1+e\cos(\phi)},
\end{equation}
with the distances of the stars from the center of mass being:
\begin{equation}
d_1 = \left( \frac{m_2}{m_1+m_2} \right)d,  \qquad d_2 = \left( \frac{m_1}{m_1+m_2} \right)d.
\end{equation}

The non-zero quadrupole moment components are:
\begin{align}
Q_{xx} = \mu d^2 \cos^2 \phi ,\notag \\
Q_{yy} = \mu d^2 \sin^2 \phi ,\notag \\
Q_{xy} = Q_{yx} = \mu d^2 \sin \phi \cos \phi \notag \\
\end{align}
with $\mu=m_1 m_2/(m_1+m_2)$.  
Note that the $Q_{ij}$ are expressed here in terms of the gravitational mass instead of the 
conserved mass $m^{\ast} = \int \rho^{\ast} d^3 x$ as used in the previous section.
Switching between the two involves corrections of order $v^2$ which do not
affect the leading order calculation of the orbital evolution.
The second and third time derivatives of the quadrupoles are 
needed for (\ref{edot1}) and (\ref{ldot3}).
Using the angular velocity: 
\begin{equation}
\omega = \frac{[(m_1+m_2)a(1-e^2)]^{1/2}}{d^2}
\end{equation}
one finds the second derivatives to be:
\begin{align}
\frac{d^2Q_{xx}}{dt^2}=-\gamma(4\cos(2\phi)+e(3\cos(\phi)+\cos(3\phi))) , \notag \\
\frac{d^2Q_{yy}}{dt^2}=\gamma(4\cos(2\phi)+e(4e+7\cos(\phi)+\cos(3\phi))) , \notag \\
\frac{d^2Q_{xy}}{dt^2}=\frac{d^2Q_{yx}}{dt^2}= 
-2\gamma \sin(\phi)(4\cos(\phi)+e(3+\cos(2\phi))) , \notag \\
\end{align}
where $\gamma$ defined as:
\begin{equation}
\gamma = \frac{m_1m_2}{2a(1-e^2)}.
\end{equation}
The third derivatives evaluate to:
\begin{align}
\frac{d^3Q_{xx}}{dt^3}=\beta(1+e\cos\phi)^2(2\sin2\phi+3e\sin\phi\cos^2\phi) , \notag \\
\frac{d^3Q_{yy}}{dt^3}=-\beta(1+e\cos\phi)^2(2\sin2\phi+e\sin\phi(1+3\cos^2\phi)) , \notag \\
\frac{d^3Q_{xy}}{dt^3}=\frac{d^3Q_{yx}}{dt^3}= \notag \\
-\beta(1+e\cos\phi)^2(2\cos2\phi-e\cos\phi(1-3\cos^2\phi)) , \notag \\
\end{align}
with $\beta$ defined as:
\begin{equation}
\beta^2 = \frac{4m^2_1m^2_2(m_1+m_2)}{a^5(1-e^2)^5}.
\end{equation}
The time derivative of the monopole $\partial_t M$ is also needed 
for eccentric binaries:
\be
\partial_t M = -\frac{1}{2} (m_1\partial_t v_1^2 + m_2\partial_t v_2^2)
=\frac{1}{2}\beta e \sin\phi(1+e\cos\phi)^2.
\ee

The energy therefore radiates at the rate:
\be
\partial_t E = - \frac{1}{15}\beta^2 (1+e\cos\phi)^4 (4+2e^2+8e\cos\phi + 2e^2\cos^2\phi),
\label{edot2}
\ee
and likewise for the angular momentum: 
\be
\partial_t L_z = - \frac{2}{15} \beta \gamma (1+e\cos\phi)^3 (8-2e^2+12e\cos\phi + 6e^2\cos^2\phi).
\label{ldot4}
\ee
In general relativity the gravitational wave 
energy can't be localized at individual points in space, and therefore 
the expressions for the $dE/dt$ and $dL_z/dt$ need to be averaged over an orbit.
While the energy can be localized in Nordstr\"{o}m's theory, we will also average 
(\ref{edot2}) and (\ref{ldot4}) in order to directly compare with general relativity.  
This is not a bad approximation since the orbital parameters 
evolve even more slowly than in GR.  
We find: 
\be
\langle\dot{E}\rangle = - \frac{16}{15}\frac{m^2_1 m^2_2 (m_1+m_2)}{a^5 (1-e^2)^{7/2}}
\left[ 1+\frac{13}{4}e^2+\frac{7}{16}e^4 \right] \label{dedt_final}
\ee
which is similar to the value given by Peters \cite{peters64} for general relativity
(differing only in the coefficients):
\be
\langle \dot{E} \rangle = - \frac{32}{5}\frac{m^2_1 m^2_2 (m_1+m_2)}{a^5 (1-e^2)^{7/2}}
\left[ 1+\frac{73}{24}e^2+\frac{37}{96}e^4 \right].
\ee
Likewise for the angular momentum we average to find:
\be
\langle \dot{L_z} \rangle = - \frac{16}{15}\frac{m^2_1 m^2_2 (m_1+m_2)^{1/2}}{a^{7/2} (1-e^2)^2}
\left[ 1+\frac{7}{8}e^2 \right],
\ee
which again is one sixth the value found in general relativity. 

Finally, by using 
\begin{align}
a= - m_1 m_2/2E, \\
L^2 = m^2_1 m^2_2 a (1-e^2)/(m_1+m_2)
\end{align}
 $\langle \dot{E} \rangle$ and $ \langle \dot{L_z} \rangle$ can be converted 
into rates of change for the semi-major axis $\langle \dot{a} \rangle$ 
and eccentricity $\langle \dot{e} \rangle$:
\be
\left\langle \frac{da}{dt} \right\rangle = -\frac{32}{15}\frac{m_1 m_2 (m_1+m_2)}{ a^3 (1-e^2)^{7/2}}
\left[ 1+\frac{13}{4}e^2+\frac{7}{16}e^4 \right] , \label{dadt}
\ee
and
\be
\left\langle \frac{de}{dt} \right\rangle = -\frac{18}{5} \frac{m_1 m_2 (m_1+m_2)}{ a^4 (1-e^2)^{5/2}}
e\left[ 1+\frac{7}{18}e^2 \right] . \label{dedt}
\ee

In the zero eccentricity case the semi-major axis integrates to:
\be
a(t)=\left( a^4(0) - \frac{128}{15} m_1 m_2 (m_1+m_2) t \right)^{1/4} . \label{inspiral}
\ee 
Reproducing this $1/4$ power law inspiral, similar to the expression found 
by Peters and Mathews for GR, is a major test of the Nordstr\"{o}m simulation.

We also simulate binaries that retain some eccentricity 
$e$.  
Equations (\ref{dadt}) and (\ref{dedt}) can be combined to directly compare how $a$ and $e$ evolve:
\be
\left\langle \frac{da}{de} \right\rangle = \frac{16}{27} \frac{a}{e (1-e^2)} 
\frac{1+\frac{13}{4}e^2 + \frac{7}{16}e^4}{1+\frac{7}{18}e^2} \label{dade} .
\ee
An eccentric Nordstr\"om binary will therefore circularize as it evolves.  
Reproducing this rate of circularization, and the rate of precession 
given in (\ref{precess_an}), are the major tests of eccentric binaries 
in our numerical simulation.

\section{Numerical Methods}

\subsection{Co-rotating Coordinates and the Weak Radiation Reaction Approximation}

The field equation (\ref{eomfield}) needs to be finite differenced to simulate a 
binary system in Nordstr\"om's theory.
We begin by rewriting the wave equation $-\partial^2_t \varphi + \nabla^2 \varphi$ 
in a corotating spherical coordinate system.
The corotating azimuthal angle $\bar{\phi}$ is given in terms of the original by: 
 $\bar{\phi} = \phi - \Omega(t)$ (with $\Omega(t) = \int_0^t \omega(t')dt'$ 
and $t$, $r$, and $\theta$ remaining unchanged) where $\omega(t)$ always 
instantaneously matches the binary's angular velocity.
The differential equation becomes (dropping the bar notation):
\begin{align}
-\partial^2_t \varphi + \dot{\omega} \partial_{\phi} \varphi 
+ 2 \omega \partial_t \partial_{\phi} \varphi - \omega^2 \partial^2_{\phi} \varphi 
+\nabla^2 \varphi = \\ \notag
4 \pi (1+\varphi)^3 (\rho(1+\varepsilon) -3p).
\end{align}

Spherical harmonics are used to split this 3+1 differential equation 
into a set of 1+1 radial equations, one for each $l$ and $m$ term:
\begin{align}
-{\partial_t}^2\varphi_{lm}+  i m \dot{\omega} \varphi_{lm}
+ 2 i m \omega \partial_t \varphi_{lm} + m^2 {\omega}^2 \varphi_{lm}  + \notag \\
\frac{\partial_r [r^2 \partial_r \varphi_{lm}]}{r^2} 
- \frac{l(l+1) \varphi_{lm}}{r^2} = 4\pi S_{lm}(t,r) , \label{tdep2}
\end{align}
where the source term is also decomposed:
\be
S_{lm}(t,r) = \int d\Omega Y^{\ast}_{lm}(\theta,\phi)[(1+\varphi)^3(\rho(1+\varepsilon)-3p)].
\label{coro1}
\ee
We have written a spherical harmonics package as part of the code in order to
evaluate (\ref{coro1}) and also synthesize the field $\varphi$
from its $\varphi_{lm}$ components.
We will use (\ref{tdep2}) to simulate binaries that retain some degree of eccentricity.

For binaries that are in quasi-circular orbits, 
we make the Weak Radiation Reaction (WRR) approximation, 
which is based on the presence of two timescales: the fast orbital motion 
characterized by $\omega$ (with period $T \sim \omega^{-1}$), and the longer radiation reaction
timescale $\tau$.  For a binary with separation $d$ the ratio of these timescales goes as 
$\tau/T \sim (d/M)^{5/2}$.  Thus there is a hierarchy of scales among the time derivatives:
\begin{equation}
\frac{\partial}{\partial t^2} \sim \frac{1}{\tau^2}, \qquad \omega\frac{\partial}{\partial t} \sim 
\dot{\omega} \sim \frac{1}{T\tau},  \qquad
\omega^2\frac{\partial}{\partial \phi^2} \sim \frac{1}{T^2} .
\end{equation}

The second time derivative term is the smallest, and is dropped when making the 
WRR approximation, giving a first
order in time differential equation (which resembles the 1-D Schr\"odinger equation):
\begin{align}
\partial_t \varphi_{lm} = \frac{i}{2 m \omega} [\frac{\partial_r [r^2 \partial_r \varphi_{lm}]}{r^2} 
+(m^2 {\omega}^2 +  i m \dot{\omega} \label{tdep1}  \notag \\ 
- \frac{l(l+1)}{r^2})\varphi_{lm}  - 4 \pi S_{lm}(t,r)] 
\end{align}
for the $m\neq0$ terms and 
\begin{align}
\frac{\partial_r [r^2 \partial_r \varphi_{l0}]}{r^2} - \frac{l(l+1)}{r^2}\varphi_{l0}  - 4 \pi S_{l0}(t,r)=0
\label{tdep0} 
\end{align}
for the time independent $m=0$ terms.

To solve for the initial data for the Cauchy evolution  
the single time derivative terms in equation (\ref{tdep1}) are dropped, giving:
\begin{align}
\frac{\partial_r [r^2 \partial_r \varphi_{lm}]}{r^2} 
+\left( m^2 {\omega}^2 - \frac{l(l+1)}{r^2} \right) \varphi_{lm} \label{tind1} \notag \\ 
- 4 \pi S_{lm}(t,r)=0 .
\end{align}

This leaves the boundary conditions.
The inner boundary is given by setting the 
radial derivative of the $\varphi_{lm}$ terms to zero:
\be
\partial_r \varphi_{lm} = 0.
\ee
At the outer 
boundary the Sommerfeld outgoing wave boundary condition is used, 
which has the following form in the co-rotating reference frame:
\be
\partial_t \varphi_{lm} = i m \omega \varphi_{lm} - (1/r)\varphi_{lm} 
-\partial_r \varphi_{lm} . \label{sommer}
\ee

The binaries are first modeled in a quasi-circular orbit 
(with the expected inspiral given by (\ref{inspiral})) 
with equations (\ref{tdep1}) and (\ref{tdep0}) used 
to evolve the field $\varphi$.
As noted equation (\ref{tdep1}) closely resembles the Schr\"odinger equation,
so it is finite differenced using the
Crank-Nicholson method (see e.g. \cite{recipes}).  

The hyperbolic equation (\ref{tdep2}) needs to be finite differenced for inspirals 
that retain some eccentricity.
Given the success we have in evolving (\ref{tdep1}) with Crank-Nicholson,
we devise a modified Crank-Nicholson for (\ref{tdep2}):
\begin{align}
\label{crank2}
-(\varphi^{N+1}_i - 2 \varphi^{N}_i + \varphi^{N-1}_i)
+  i m \omega \Delta t (\varphi^{N+1}_i - \varphi^{N-1}_i)  = \\ \notag
\frac{\Delta t^2}{2}
(-\frac{(\varphi^{N+1}_{i+1} - 2 \varphi^{N+1}_i + \varphi^{N+1}_{i-1})}{\Delta r^2} 
-\frac{(\varphi^{N+1}_{i+1} - \varphi^{N+1}_{i-1})}{r \Delta r}  \\ \notag
+  (\frac{l(l+1)}{r^2} - m^2 \omega^2 - i m \dot{\omega})\varphi^{N+1}_i )  \\ \notag
+ \frac{\Delta t^2}{2}
(-\frac{(\varphi^{N-1}_{i+1} - 2 \varphi^{N-1}_i + \varphi^{N-1}_{i-1})}{\Delta r^2} 
-\frac{(\varphi^{N-1}_{i+1} - \varphi^{N-1}_{i-1})}{r \Delta r} \\ \notag
+(\frac{l(l+1)}{r^2} - m^2 \omega^2 - i m \dot{\omega})\varphi^{N-1}_i ) \\ \notag
+ 2\pi \Delta t^2 (S^{N+1}_i + S^{N-1}_i) .
\end{align}
We find that this balanced implicit method also allows for fast, stable, and accurate 
field evolutions.

\subsection{Numerical Tests}

\begin{figure}
\includegraphics[width=3.5in]{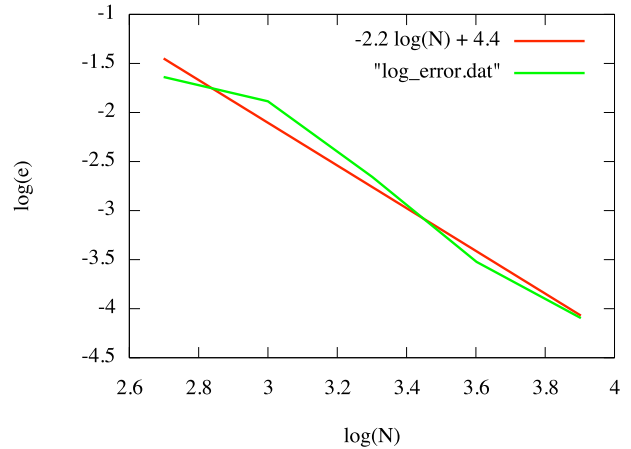}
\caption{\label{newt_conv} Plotted is the log${}_{10}$ of the error $e=\sup |\varphi_{Newton}/\varphi-1|$ 
against the log${}_{10}$ of the number of radial grid points $N$.  
A linear fit gives a slope of $-2.2 \pm 0.2$, i.e. second order convergence.  
The lowest resolution grid uses $N=500$ radial grid-points and 
$L$ and $M$ values up to $25$.  This is then doubled four times up to 
resolution with $N=8000$ radial grid points and $L$ and $M$ values up to $400$.}
\end{figure}

We use a mixture of finite difference and pseudo-spectral methods, and these 
need to be tested first before evolving the full system.
A useful test case is provided by the Newtonian system, 
with the inhomogeneous wave equation reducing to 
Poisson's equation:
\be
\nabla^2 \varphi = 4 \pi \rho \label{newt1} ,
\ee
and the acceleration is given by $F=ma$:
\be
\frac{d^2 x^i_a}{dt^2} = \frac{1}{M_a} \int_a \rho \partial_i \varphi d^3 x \label{newt2}.
\ee
Equation (\ref{newt1}) can be split into its spherical harmonic components by dropping 
the $m^2\omega^2\varphi_{lm}$ term from equation (\ref{tind1}): 
\be
\frac{\partial_r [r^2 \partial_r \varphi_{lm}]}{r^2} - \frac{l(l+1)}{r^2} \varphi_{lm} 
+ 4 \pi S_{lm}(t,r)=0 , \label{newt3}
\ee
with the source terms given by $S_{lm}(t,r) = \int d\Omega Y^{\ast}_{lm}(\theta,\phi)\rho(t,r,\theta,\phi)$.
Equation (\ref{newt3}) is then finite differenced in the standard second order manner.
In order to check the convergence of the numerical solution 
we choose a matter distribution for the stars given by $\rho(r) = \rho_0(1-r^2/R^2)$ 
(with some central density $\rho_0$ and radius $R$) 
which allows for an exact analytical solution for $\varphi$.  
After solving for the $\varphi_{lm}$ components based on this source
and then synthesizing the field $\varphi$,
we find that our numerical solution converges at second order to the analytical solution, 
as shown in figure (\ref{newt_conv}).


The Newtonian system presents a second test in addition to the convergence of 
the field: the integral in equation (\ref{newt2}) must also converge to the expected 
inverse square law.
Note also that we are already using a ``hydro-without-hydro" type of approximation: the 
matter is described by a rigid density profile, so that only the acceleration of the 
center of mass needs to be determined.
We find that accurately resolving this acceleration is somewhat involved 
in a spherical coordinate system.  For a given resolution, and using 
second order methods, we generally find that the solution for the field (\ref{newt1}) 
is much more accurate than the result for the acceleration (\ref{newt2}).
The error in the computed acceleration also grows quite quickly as the separation 
increases. 
The error does converge to 
zero as the resolution is increased, but inconveniently high 
resolutions are needed to reduce it to acceptable values.  

This inconvenience arises because (\ref{newt2}) includes the 
gradient of the star's self field. 
The Newtonian self force integrates to zero 
analytically, but numerically there will be some small residue.
Furthermore, the amplitude of the self field is much larger the $1/r$ contribution from the other 
star, so a small percent error in calculating its gradient can swamp 
the correct $1/r^2$ contribution from the other star, especially as the 
separation increases and the resolution inside the stars decreases (for a fixed spherical grid).

\begin{figure}
\includegraphics[width=3.5in]{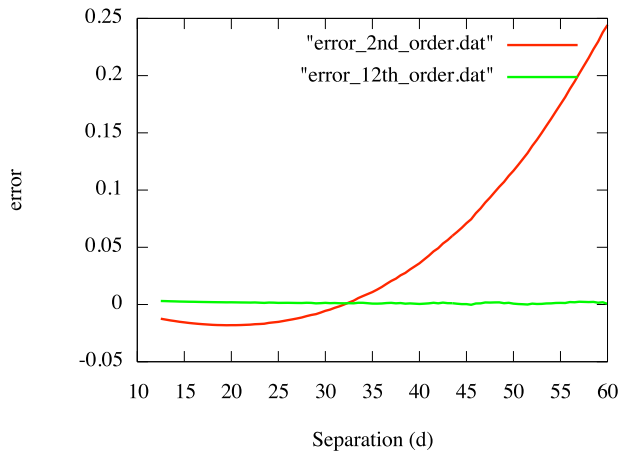}
\caption{\label{ar_e2_e12}Plotted is the error = $(a_r)/(1/d^2)-1$ of the radial acceleration
(compared to the expected Newtonian inverse square law with a mass of $1$) 
as a function of the separation $d$ for the 2nd order and 12th order derivative methods.  
The standard 2nd order accurate method gives an error 
of 25 percent for a separation of $60 M$ 
(for a particular grid resolution),
while the 12th order method gives an error of $\sim$ 0.3 percent, 
which is the same accuracy the field is resolved to at this separation.}
\end{figure}

Experimentation shows that the error grows almost as quickly 
if the analytic solution for $\varphi$ is inserted into the mesh and then 
finite differenced to calculate (\ref{newt2}).  However, if the analytical derivatives 
$\partial_i \varphi$ are inserted into (\ref{newt2}) then the result for the acceleration 
is much more accurate -- now comparable to the accuracy of (\ref{newt1}).
We thus need much more accurate derivatives, which are available through  
higher order methods.  
Experimentally, each increase in the order 
of the derivatives gave better results (as compared to the analytical 
solutions), so we used 12th order accurate derivatives, given by:
\begin{align}
\partial_i \varphi \simeq \frac{1}{27720 \Delta x}[5\varphi_{i-6} - 72\varphi_{i-5}
+495\varphi_{i-4} \notag 
\\ -2200\varphi_{i-3}+7425\varphi_{i-2}-23760\varphi_{i-1} \notag \\
+23760\varphi_{i+1}-7425\varphi_{i+2}+ 2200\varphi_{i+3} \notag \\
-495\varphi_{i+4}  + 72\varphi_{i+5} - 5\varphi_{i+6}] .
\end{align}
The resulting improvement in the acceleration is shown in 
figure (\ref{ar_e2_e12}).  
Note that we do not claim to have a 12th order accurate
code, as the field $\varphi$ is still solved for with 2nd order 
accurate implicit methods.  
However, these higher order derivatives do allow the 
net accelerations computed in (\ref{newt2}) to have the same level of accuracy 
as the solution for $\varphi$.

\begin{figure}
\includegraphics[width=3.5in]{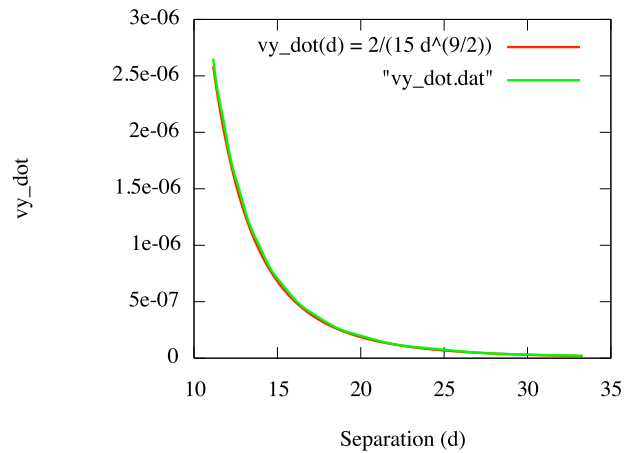}
\caption{\label{aphi_plot}Plot of the measured 
acceleration $d v^y/dt$ (i.e. in the $\phi$ direction) as a 
function of separation, which closely matches 
the expectation given in \ref{aphi_92}.}
\end{figure}

The next test is to promote the Laplacian operator in equation (\ref{newt1})
to a wave operator:
\be
\Box \varphi = 4 \pi \rho . \label{box1}
\ee
The spherical harmonic decomposition for this in corotating coordinates 
was given in (\ref{tdep2}) and the initial data is found by solving (\ref{tind1}).  
The homogeneous version of  (\ref{tind1})
is solved by the Bessel and Neumann functions
$j_l(m \omega r)$ and $\eta_l (m \omega r)$, or in the case of the Sommerfeld 
boundary condition (\ref{sommer}), by an outgoing wave spherical Hankel function.  
At large $r$ this asymptotically approaches a sinusoidal function of frequency 
$m\omega$ with a $1/r$ envelope.  
Our finite difference code correctly reproduces this asymptotic behavior.  


This simple wave equation system also provides another test 
of the net acceleration of the center of mass.  
For this test we retain the Newtonian equation (\ref{newt2}) to calculate accelerations
based on the field $\varphi$.  In Newtonian language, 
the wave equation $\Box \varphi$ gives rise to
a radiation reaction potential $\Phi^{(react)}$ 
(when driven by the orbital motion of the compact bodies).  
This potential $\Phi^{(react)}$ results in a drag force acting on the stars,
causing the orbit to decay in a manner that closely matches the full theory.  
A quadrupole calculation similar to the full Nordstr\"om theory version 
finds that the orbit decays at the rate: 
$\dot{d} = -(64/15)m_a^3/d^3$ where $d$ is the separation and 
$m_a=m_b$ is the mass of the equal mass stars.
Let the two stars be instantaneously 
aligned along the x-axis in a Cartesian plane, at $\pm d/2$
away from the origin, with Newtonian velocities $v^y = \pm \hat{y} (m_a/2d)^{1/2}$.
The gradient of $\Phi^{(react)}$ then gives an acceleration 
opposite to the velocity (i.e., in the $\phi$ direction):
\be
\dot{v}^y=(1/15)m_a^{7/2}/(d/2)^{9/2}, \label{aphi_92}
\ee
thereby causing a quasi-circular inspiral.  
The code accurately 
reproduces the inverse nine-halves scaling of this acceleration, 
as shown in Figure (\ref{aphi_plot}), where we have set $m_a=1/2$ 
(so that the total mass is $1$).

\section{Modeling Nordstr\"{o}m's Theory}

With the initial tests passed we can now simulate Nordstr\"{o}m theory.  
Again we solve for the initial data with (\ref{tind1}), but we do not decompose
the source as we did in (\ref{coro1}).  Instead we rewrite the source term 
using the conserved density $\rho^{\ast}$, finding:
\be
S_{lm}(t,r) = \int d\Omega Y^{\ast}_{lm}(\theta,\phi)[\sqrt{1-v^2}\rho^{\ast}(1+\varepsilon-3p/\rho)] .
\label{decomp2}
\ee
This removes much of the nonlinearity from the equation of motion for 
the field.  
We find $\varepsilon$ and $-3p/\rho$ using $\rho^{\ast}$ (which is held 
constant under the hydro-without-hydro approximation), and the value 
of $\varphi$.

We first model quasi-circular inspirals 
by making use of the WRR approximation.
Initial data is found by starting with a circular Keplerian 
binary, and then iteratively tuning the orbital parameters to 
reduce the initial eccentricity.
The field is then evolved 
forward in time via (\ref{tdep1}), which is finite differenced with the Crank-Nicholson scheme.  
Note that it is crucial to also time average the source $S_{lm}(t,r)$ 
in the Crank-Nicholson scheme, as otherwise the source and field will be a 
half time step out of sync which results in spurious accelerations.  

\begin{figure}
\includegraphics[width=3.5in]{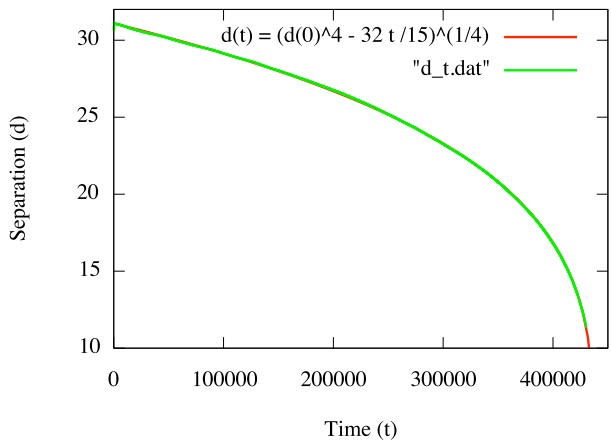}
\caption{\label{inspiral_plot}Separation $d$ as a function of time for a quasi-circular inspiral.
The inspiral closely matches the 1/4 power law result given in (\ref{inspiral}).}
\end{figure}

\begin{figure}
\includegraphics[width=3.5in]{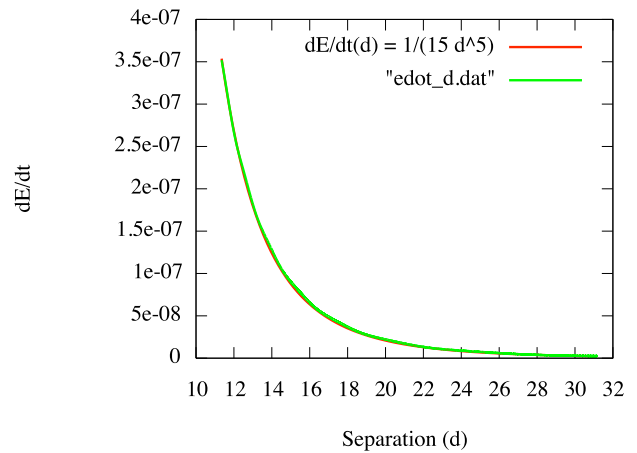}
\caption{\label{eloss_plot}Plot of the rate of energy loss of the binary $dE/dt$ as a function 
of the separation $d$, compared to the theoretical value given in (\ref{dedt_final}), 
where $m1=m2=0.5$ and $e=0$.}
\end{figure}

\begin{figure}
\includegraphics[width=3.5in]{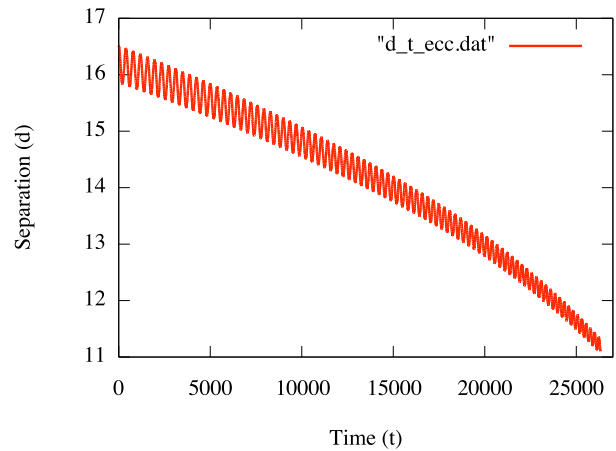}
\caption{\label{d_t_ecc} 
Separation $d$ as a function of time for a binary with an 
initial eccentricity of $e \sim 0.021$.}
\end{figure}

\begin{figure}
\includegraphics[width=3.5in]{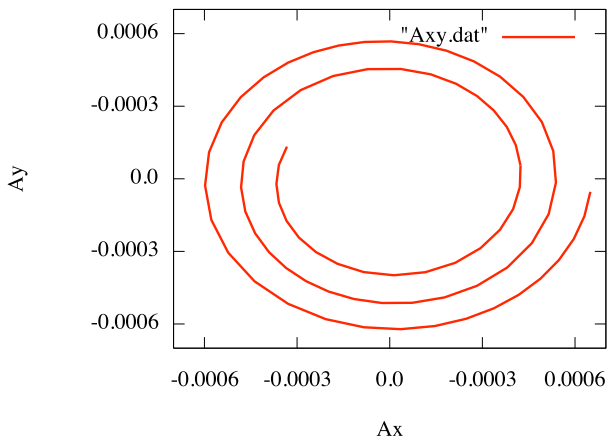}
\caption{\label{runge_lenz} 
Plot of the Runge Lenz vector $\bold{A}$ as it evolves, 
which demonstrates that
binary precesses about 2.5 times over the course of the
inspiral.  The circularization of the binary can also be seen 
as the magnitude of $\bold{A}$ drops by about a factor 
of two.}
\end{figure}

\begin{figure}
\includegraphics[width=3.5in]{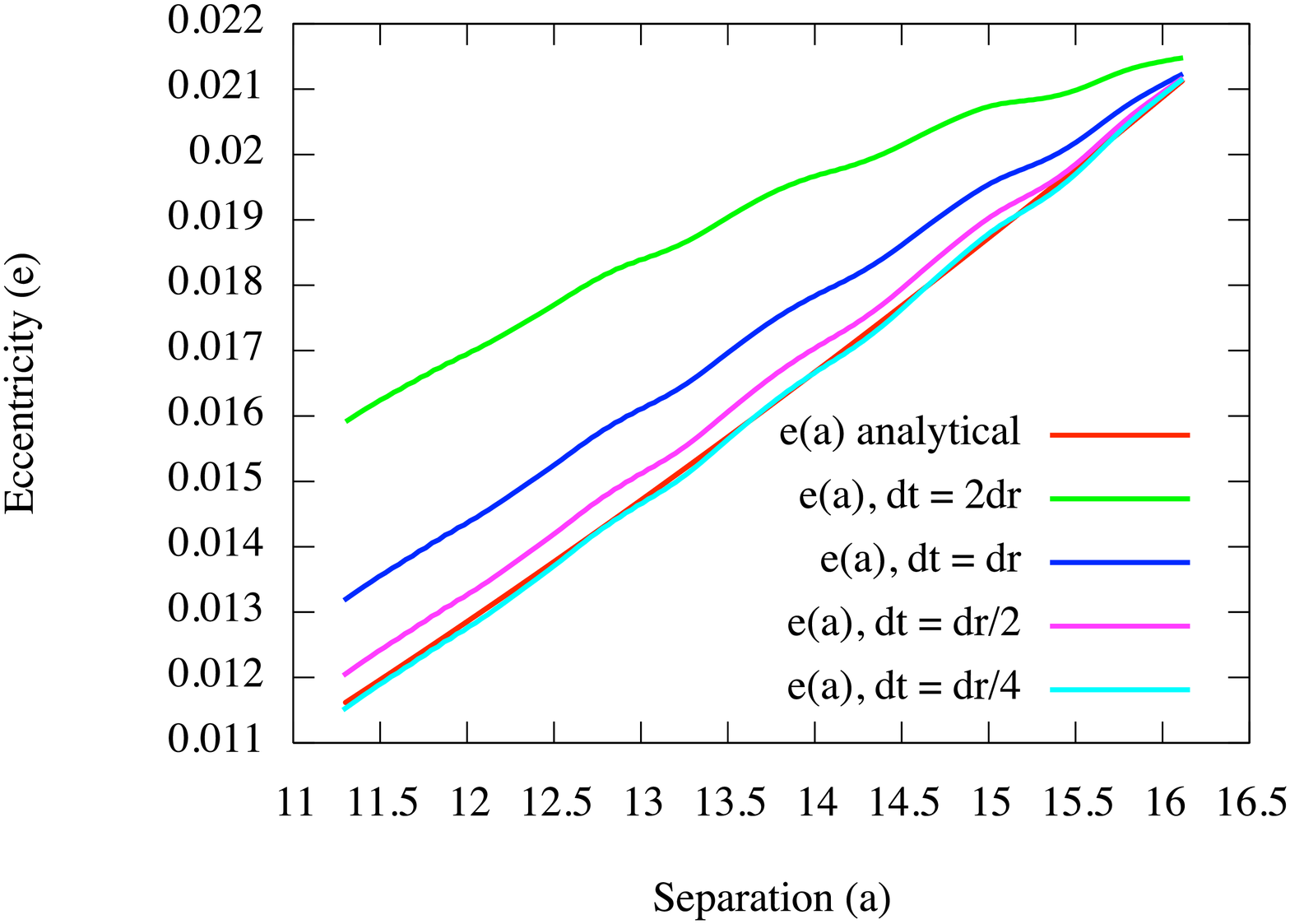}
\caption{\label{a_vs_e} 
Plot of the eccentricity $e$ as a function of the semi-major axis $a$
for simulations with a time step $\Delta t$ ranging from 2.0 to 0.25 $\Delta r$ vs. the 
analytical expectation from (\ref{dade}).}
\end{figure}

After each time step the field $\varphi(r,\theta,\phi)$ is synthesized from its 
spherical harmonic components and used to solve for the acceleration using a 
discretized combination of (\ref{bseom2}) and (\ref{bseom3}).
As noted, 12th order finite differencing is used for the 
derivatives of $\varphi$ so that the net acceleration is accurate.
The components of equation (\ref{bseom3}) are also expanded out in the corotating 
reference frame.
The corotating coordinate transformation adds shift vector like terms
to the metric which in turn leads to new terms in the 
sum of connection coefficients $\Gamma^j_{\alpha\beta}v^{\alpha}v^{\beta}$.  
These additions effectively add a centripetal force, 
so that when the correct $\omega$ is found (via iterative testing) 
the stars will be nearly at
rest in the co-rotating frame (expect for the slow merger due to the 
radiation reaction).  

The masses of the stars are  
set to $m_a = m_b = 1/2$ 
for both the quasi-circular and 
eccentric simulations. 
An initial central density of $\rho^{\ast}_c \sim 3.72 \times 10^{-4}$ and polytropic parameters 
$\Gamma=5/3$ and $\kappa=7.29$ are picked to 
give a compactness of $m_a/R_a = 0.1$.   
The stars are then rescaled in the simulation to give a total 
mass of 1, so that each star has a radius of $5 M$.
The outer radial boundary is set to $r_{outer} = 560 M$, 
which places it in the wave zone 
for an initial separation of $31 M$.
The radial grid spacing is $\Delta r=0.4 M$, 
and spherical harmonic modes up to 
$l=m=80$ are used 
(in general the discretization is chosen so that 
there is comparable resolution in 
the angular and radial directions).
As we use Crank-Nicholson to evolve (\ref{tdep1}) for the 
quasi-circular case 
we can utilize a time step of $\Delta t=6.0 M$.
This is much larger than the value given by the  
Courant-Friedrichs-Lewy (CFL) condition,
but is sufficient to resolve the slow evolution 
of the system on the radiation reaction timescale.
A large time step also proves to be stable for eccentric binaries, 
but in this case there is dynamical motion on the shorter orbital time scale, 
and experimentation shows that 
time steps similar to the CFL condition are needed in order to get 
accurate results.

The binary separation as a function of time for 
the quasi-circular evolution 
is given in Figure (\ref{inspiral_plot}).
The initial separation is set to be $31 M$ and the stars 
then inspiral to a separation of $11 M$ (i.e. almost to merger)  
over the course of $\sim 600$ orbits.  
The inspiral closely 
matches the $1/4$ power law (\ref{inspiral}) 
predicted by the semi-analytical calculation.  
In Figure (\ref{eloss_plot}) we plot the rate of energy loss  
as measured directly at the outer shell of the domain.
We find a close match to the analytical prediction given in (\ref{dedt_final}).  
The energy contained in 
the volume of the computational domain is also tracked 
using (\ref{energy}) and found 
to decrease in sync with the energy flux through the outer boundary.  
The quasi-circular inspiral therefore closely matches the analytical expectations.

Next we examine binaries 
that retain a small amount of eccentricity.  
The WRR approximation is dropped and 
the field is evolved with the modified 
Crank-Nicholson scheme given in (\ref{crank2}).
We select a value of $\omega$ which gives the 
binary a small amount of eccentricity ($e \sim 0.02$)  
at an initial separation of $a \sim 16 M$.
With this small value of $e$ the binary decays only slightly faster 
than a quasi-circular binary with the same 
semi-major axis.  
The system circularizes noticeably over the evolution 
(finishing with $e \sim 0.01$) 
as can be seen in Fig. (\ref{d_t_ecc}).

We first compare the rate of precession to the analytical 
profile (\ref{precess_an}) given by Post Newtonian calculations.
We find that the code matches this
expectation closely: 
for the chosen initial data the binary precesses 
about $5 \pi$ radians over the course of dozens of orbits 
as the separation 
$d$ decreases from $16 M$ to $11 M$.
This precession can be 
seen by fitting the Runge-Lenz vector $\bold{A}$
to the binary's orbital parameters
(and then averaging over an orbit to remove 
any orbital timescale artifacts).
$\bold{A}$ points along the semi-major axis, 
and is proportional to the eccentricity.
It therefore compactly displays both the precession 
and circularization of a binary,
as shown in Figure (\ref{runge_lenz}).

The gradual circularization of the binary 
can also be seen in Figure (\ref{a_vs_e}), where 
an extracted value for  
$e(a)$ (for simulations with a range of time steps) is compared 
to the result found by integrating the analytical 
form in (\ref{dade}).  
The slow secular evolution of $e$ is quite sensitive to accumulating 
errors, and we find that the time step needs to be reduced to 
$\Delta t \sim \Delta r/4$ in order to get accurate results.

\section{Conclusions}

The numerical techniques we have developed work well in simulating 
binary inspirals in Nordstr\"om's theory.  
The co-rotating spherical coordinate system is 
quite sensitive to the 2.5 PN order radiation reaction force, 
and is well suited to conserving angular momentum, allowing the for 
accurate evolution of inspirals over many orbits
($\sim 600$ in the case of the quasi-circular inspiral).
The code demonstrates this, as it accurately matches
the analytic predictions we made for Nordstr\"om's theory, 
from the 1/4 power law profile for a quasi-circular inspiral 
and the corresponding rate of energy loss, 
to the rate of precession and circularization 
of an eccentric binary.

As a side benefit we find that Nordstr\"om's second theory is a useful laboratory 
for developing computational techniques for numerical relativity.  
It is a fully relativistic metric theory, and yet has fairly simple field equations
which can be put in a nearly linear form.  
Furthermore the theory follows the SEP and thus 
stars move on Keplerian 
orbits to lowest order, 
and there are no star-crushing effects.

Our main goal is to use our corotating spherical coordinates framework 
to model the late inspirals of binaries in GR.
We explore a modification of a generalized harmonic formulation of the 
Einstein equations to this end in \cite{garrett09}.  
Given the success we 
have had in applying it to Nordstr\"om binaries, we feel confident 
that it should allow for fast, long and accurate simulations in GR as well.

 
\bibliographystyle{unsrt}
\bibliography{analytical}

\end{document}